%% file: converse6m-arxive.tex
\documentclass[11pt, a4paper]{article}

\author{Matthias Christandl \thanks{Centre for Quantum Computation, DAMTP, University of Cambridge,
Wilberforce Road, Cambridge, CB3 0WA, UK,
matthias.christandl@qubit.org} \and Aram W.
Harrow\thanks{Department of Computer Science, University of
Bristol, Bristol, BS8 1UB, UK, a.harrow@bris.ac.uk} \and Graeme
Mitchison\thanks{Centre for Quantum Computation, DAMTP, University
of Cambridge, Wilberforce Road, Cambridge, CB3 0WA, UK,
g.j.mitchison@damtp.cam.ac.uk}}

\date{\today}

\include{conversedef}

\title{On Nonzero Kronecker Coefficients and their Consequences for Spectra}
\begin{document}
\maketitle

\begin{abstract}
A triple of spectra $(r^A, r^B, r^{AB})$ is said to be
\emph{admissible} if there is a density operator $\rho^{AB}$ with
$$ (\spec \rho^A, \spec \rho^B, \spec \rho^{AB})=(r^A, r^B,
r^{AB}).$$ How can we characterise such triples? It turns out that
the admissible spectral triples correspond to Young diagrams
$(\mu, \nu, \lambda)$ with nonzero Kronecker coefficient $g_{\mu
\nu \lambda}$~\cite{ChrMit05, Klyachko04}. This means that the
irreducible representation $V_\lambda$ is contained in the tensor
product of $V_\mu$ and $V_\nu$. Here, we show that such triples
form a finitely generated semigroup, thereby resolving a
conjecture of Klyachko~\cite{Klyachko04}. As a consequence we are
able to obtain stronger results than in~\cite{ChrMit05} and give a
complete information-theoretic proof of the correspondence between
triples of spectra and representations. Finally, we show that
spectral triples form a convex polytope.
\end{abstract}

\section{Introduction}

A curious connection between representation theory and the spectra
of operators was discovered recently. Suppose we are given a
bipartite density operator $\rho^{AB}$, and suppose this has
spectrum $r_{AB}=\spec(\rho^{AB})$. Let $r^A$ be the spectrum of
the marginal operator $\rho^A=\tr_B \rho^{AB}$, and $r^B$ that of
the other marginal operator $\rho^B$. Then clearly there are
restrictions on the possible spectral triples $(r^A, r^B, r^{AB})$
as $\rho^{AB}$ ranges over all density operators. For instance, if
$\rho^{AB}$ is pure, so $r^{AB}=(1,0,...)$, then $r^A=r^B$. How
does one characterise the set of possible spectral triples? One
way to do this is via representation theory~\cite{ChrMit05,
Klyachko04}: there is a correspondence between triples of spectra
and irreducible representations of the symmetric group $V_\mu$,
$V_\nu$ and $V_\lambda$, where \be \label{eq-kron} V_\lambda
\subset V_\mu \otimes V_\nu.\ee

Two rather different methods were used to prove this.
In~\cite{Klyachko04} a body of powerful techniques from invariant
theory~\cite{Heckman82, MuFoKi94, BerSja00} were harnessed.
In~\cite{ChrMit05}, the approach came from the direction of
quantum information theory, and a key ingredient was a theorem
relating spectra and Young diagrams due to Alicki, Rudnicki and
Sadowski~\cite{AlRuSa87} and Keyl and Werner~\cite{KeyWer01PRA}.
The latter theorem can be given a short and elegant
proof~\cite{HayMat02} (see also~\cite{ChrMit05}) that has
interesting parallels with classical information theory. To those
with an information theory background, therefore, the approach
taken in~\cite{ChrMit05} has some advantages of accessibility. It
is shown there that for every density operator $\rho^{AB}$ there
is a sequence of triples $(\mu^{(j)}, \nu^{(j)}, \lambda^{(j)})$
satisfying relation~(\ref{eq-kron}) that converges to the spectra:
$$
\lim_{j \rightarrow \infty}(\bar\mu^{(j)}, \bar\nu^{(j)},
\bar\lambda^{(j)})=(r^A, r^B, r^{AB}),
$$
where the bar denotes normalisation. Klyachko~\cite{Klyachko04}
proves this as well as a converse that says that to every $(\mu,
\nu, \lambda)$ with $V_\lambda \subset V_\mu \otimes V_\nu$ there
is a density operator with spectra $(\bar\mu, \bar\nu,
\bar\lambda)$.

One aim of this paper is to show that informational methods can be
used to prove Klyachko's converse. On our way to this result we
prove his conjecture~\cite[Conjecture 7.1.4]{Klyachko04} that
triples $(\mu, \nu, \lambda)$ with $V_\lambda \subset V_\mu
\otimes V_\nu$ form a semigroup. We also prove that the semigroup
is finitely generated. Together with our previous results on the
correspondence with spectral triples this will imply that the set
of admissible spectral triples is a convex polytope.

\section{Background}

Let us consider in more detail the relation between irreducible
representations and spectra. The irreducible representations of
both unitary and symmetric groups are labelled by Young diagrams.
If $\lambda$ denotes a Young diagram, its row lengths are
$\lambda_1 \geq \lambda_2 \geq \ldots \geq \lambda_d$ and its size
is $|\lambda|:=\sum_{i=1}^d \lambda_i$. We denote the
corresponding irreducible representations of $\U(d)$ (or $\GL(d)$)
with highest weight $\lambda$ by $U^d_\lambda$ and those of the
symmetric group $S_k$ by $V_\lambda$. Schur-Weyl duality states
that $(\complex^d)^{\otimes k}$ decomposes as a direct sum of
irreducible representations:
\be \label{SWD} (\complex^d)^{\otimes k}\cong
\bigoplus_{\lambda\in\Par(k,d)}
U^d_\lambda \otimes V_\lambda,\ee
where $\Par(k,d)$ indicates the set of partitions of $k$ into $\leq d$
parts; i.e. the Young diagrams with no more than
$d$ rows and a total of $k$ boxes.

Consider a density operator $\rho$ on $\complex^d$. We can take
$k$ copies of it and measure the label $\lambda$ on
$(\complex^d)^{\otimes k}$. The estimation theorem~\cite{AlRuSa87,
KeyWer01PRA} states that, as $k$ increases, the spectrum $r$ of
$\rho$ is increasingly well approximated by the normalised row
lengths of $\lambda$, i.e. by the distribution
$\bar\lambda=\lambda/|\lambda|$. Formally:
\begin{theorem}[Estimation
Theorem]\label{KW}\label{theorem-Keyl-Werner} Let $P_\lambda$ be
the projection onto $U^d_\lambda \otimes V_\lambda$. Then
\be
\label{eq-KeylWerner-1}
\tr P_\lambda
\rho^{\otimes k} \leq (k+1)^{d(d-1)/2} \exp \left(-k D
(\bar{\lambda}||r)\right)\ee
where $D(p ||q)=\sum_i p_i\log(p_i/q_i)$ is the Kullback-Leibler
distance.
\end{theorem}

Let us now return to the case of bipartite states, and consider the
Clebsch-Gordan series for the symmetric group:
\[ V_\mu \otimes V_\nu \cong
\bigoplus_{\lambda\in\Par(k,k)} g_{\mu \nu \lambda} V_\lambda, \]
where the multiplicities $g_{\mu \nu \lambda}$ are known as the
{\em Kronecker coefficients (of the symmetric group)}. Since
$V_\lambda \cong V_\lambda^\star$, the Kronecker coefficients can
also be defined in terms of the $S_k$-invariant subspace
of $V_\mu \otimes V_\nu \otimes V_\lambda$,
i.e.
\be \label{kron-S-invariant} g_{\mu \nu \lambda} =\dim (V_\mu \otimes
V_\nu \otimes V_\lambda)^{S_k}. \ee

There is also a way of viewing the Kronecker coefficients in terms
of the irreducible representations of $\GL(mn)$, for integers $m,n$
satisfying $m\geq|\mu|$ and $n\geq|\nu|$.  It is arrived
at by equating the Schur-Weyl decompositions of $(\bbC^m\ot
\bbC^n)^{\ot k}$ and of $(\bbC^{mn})^{\ot k}$ (see~\cite{ChrMit05})
and reads
$$U^{mn}_\lambda\downarrow_{\GL(m)\times \GL(n)}\cong
\bigoplus_{\mu\in\bbZ_{++}^m}\bigoplus_{\nu\in\bbZ_{++}^n}
 g_{\mu \nu \lambda} U^m_\mu \otimes U^n_\nu,$$
where $\bbZ_{++}^d := \{\lambda\in\bbZ^d : \lambda_1\geq \ldots
\lambda_d \geq 0\}$ is the set of {\em dominant positive weights} for
$\GL(d)$.  This interpretation of the Kronecker coefficients can
equivalently be stated in terms of invariants as
\be
\label{kron-GL-invariant} g_{\mu \nu \lambda} =\dim (U_\mu \otimes
U_\nu \otimes U_\lambda^\star)^{\GL(m) \times \GL(n) },
\ee
 where $\GL(m)\times \GL(n)$ acts on $U^m_\mu \otimes U^n_\nu$ and
simultaneously on $U^\star_\lambda$ according to the inclusion
$\GL(m)\times \GL(n) \ra \GL(m)\otimes \GL(n) \subset \GL(mn)$.

In~\cite{ChrMit05} theorem~\ref{KW} was applied to give the
following:

\begin{theorem}\label{ourtheorem}
For every density operator $\rho^{AB}$, there is a sequence
$(\mu^{(j)}, \nu^{(j)}, \lambda^{(j)})$ of partitions, labeled by
natural numbers $j$, such that
\[g_{\mu^{(j)}, \nu^{(j)}, \lambda^{(j)}} \neq 0 \quad \mbox{ for all }\ j\]
and
    \bea
        \lim_{j \to \infty} \bar\mu^{(j)}&=&\spec \rho^{A}\\
        \lim_{j \to \infty} \bar\nu^{(j)}&=&\spec \rho^{B}\\
        \lim_{j \to \infty} \bar\lambda^{(j)}&=&\spec \rho^{AB}
    \eea
\end{theorem}
Klyachko derived a very similar theorem:

\begin{theorem} \label{theorem-Klyachko} \label{theorem-klyachko-our}
For a density operator $\rho^{AB}$ with the rational spectral
triple $(r^A, r^B, r^{AB})$ there is an integer $m>0$ such that
$g_{m r^A, m r^B, m r^{AB}}\neq 0$.
\end{theorem}
He also supplied the following converse:

\begin{theorem} \label{theorem-converse}
Let $\mu, \nu$ and $\lambda$ be diagrams with $k$ boxes and at
most $m$, $n$ and $mn$ rows, respectively. If $g_{\mu \nu \lambda}
\neq 0$, then there exists a density operator $\rho^{AB}$ on
$\cH_A \otimes \cH_B=\complex^{m} \otimes \complex^n$ with spectra
\bea \spec \rho^{A}&=&\bar{\mu}\\
\spec \rho^{B}&=&\bar{\nu}\\
\spec \rho^{AB}&=&\bar{\lambda}.
\eea
\end{theorem}
We now give a resum{\'e} of our new results.

\section{Summary of the Results}

Let $\KRON$ denote the set of triples $(\mu, \nu, \lambda)$ with
nonzero Kronecker coefficients. Our first result is

\begin{theorem}\label{theorem-stability}
$\KRON$ is a semigroup with respect to row-wise addition,
i.e.~$g_{\mu \nu \lambda} \neq 0$ and $g_{\mu' \nu' \lambda'} \neq
0$ implies $g_{\mu+\mu', \nu+\nu', \lambda+\lambda'} \neq 0$.
\end{theorem}

This was conjectured in Klyachko's paper~\cite[conjecture
  7.1.4]{Klyachko04}. It implies stability of the Kronecker
coefficients: i.e. if $g_{\mu \nu \lambda} \neq 0$ then $g_{N\mu N\nu
  N\lambda} \neq 0$, for integers $N>0$. This was announced by
Kirillov~\cite[theorem 2.11]{Kirillov04} but without proof. A
simple corollary of stability is that non-vanishing Kronecker
coefficients obey entropic relations (as explained in
\cite{ChrMit05}). More importantly, it plays a key role in giving
an information-theoretic proof of theorem~\ref{theorem-converse}.
We also present a compact version of the proof of
theorem~\ref{ourtheorem}, which was presented in~\cite{ChrMit05}.
In this way we obtain a simple proof for the full correspondence
between Kronecker coefficients and admissible spectral triples.

As a third result we will show that

\begin{theorem} \label{theorem-finitely-generated}
The semigroup $\KRON$ is finitely generated.
\end{theorem}

From this, a straightforward argument shows that

\begin{corollary}\label{cor-equiv}
Theorem~\ref{ourtheorem} and theorem~\ref{theorem-klyachko-our}
are equivalent.
\end{corollary}

Using the correspondences to spectral triples, the fact that
$\KRON$ is a finitely generated semigroup can be given the
following geometrical interpretation.

\begin{theorem} \label{theorem-convex-polytope}
$\qmp$, the set of admissible spectral triples, is a convex
polytope.
\end{theorem}

\section{The Nonzero Kronecker Coefficients form a \newline Finitely Generated Semigroup}

\noindent \fussy In order to prove
theorem~\ref{theorem-stability}, we introduce a representation of
$\GL(n)$ known sometimes as the Schwinger representation, or the
\emph{Cartan product ring}:
\be Q^n := \bigoplus_{\lambda\in\bbZ^n_{++}} U^n_\lambda.\ee
We assume here that $\lambda_n\geq 0$ because we are ultimately
interested in combining irreducible representations of $S_k$,
which are only defined for nonnegative $\lambda$.  However, all of
our results can be easily generalized for dominant weights
$\lambda$ without the restriction $\lambda_n \geq 0$.

To establish $Q^n$ as a graded ring, we introduce the {\em Cartan
product}~\cite{FultonHarris91} that maps $U_\mu\otimes U_\nu$ to
$U_{\mu+\nu}$ by projecting onto the unique $U_{\mu+\nu}$-isotypic
subspace of $U_\mu\otimes U_\nu$.  We denote the Cartan product by
$\circ$, so that for $\ket{u_\mu}\in U_\mu, \ket{u_\nu}\in U_\nu$,
$\ket{u_\mu}\circ\ket{u_\nu}$ is defined to be the projection of
$\ket{u_\mu}\ot\ket{u_\nu}$ onto the $U_{\mu+\nu}$-isotypic
subspace of $U_\mu\otimes U_\nu$.  Clearly $Q^n$ is graded under
the action of $\circ$, $\GL(n)$ preserves this grading and
$\GL(n)$ acts properly on products, i.e. $g(\ket{v}\circ\ket{w}) =
(g\ket{v})\circ(g\ket{w})$.

The proof of theorem~\ref{theorem-stability} now rests on the
following lemma:
\begin{lemma}\label{lemma:zero-div}
\begin{enumerate}
\item[(a)]
$Q^n$ has no zero divisors.  That is, if $\ket{v},\ket{w}\in Q^n$
are nonzero, then $\ket{v}\circ\ket{w}\neq 0$.
\item[(b)]
$Q^m \ot Q^n \ot (Q^{mn})^\star$ has no zero divisors.
\end{enumerate}
Here we have defined $(Q^n)^\star=\bigoplus_\lambda
(U^n_\lambda)^\star$ with corresponding Cartan product
$(U^n_\mu)^\star \circ (U^n_\nu)^\star \ra (U^n_{\mu+\nu})^\star$
and we have extended the Cartan product to tensor products in the
natural way.
\end{lemma}

\begin{proof}
Although only statement (b) of the lemma is used in the proof of
the theorem, for ease of exposition we will prove part (a) and
then sketch how similar arguments can establish (b).  Our proof is
based on the Borel-Weil
theorem~\cite[p.~115]{CarterSegalMacDonald95}, though the
presentation here is mostly self-contained.

Let $\ket{v_\lambda}$ be a highest weight vector for $U_\lambda$.
For any $\ket{\alpha}\in U_\lambda$, note that
$\bra{v_\lambda}g\ket{\alpha}$ is
\begin{enumerate}
\item[(a)] a polynomial in the matrix elements of $g$.
\item[(b)] identically zero  only if $\ket{\alpha}=0$ (due to
the irreducibility of $U_\lambda$).
\end{enumerate}
Now define the set $X_\alpha:=\{g\in\GL(n) |
\bra{v_\lambda}g\ket{\alpha} = 0\}$.  The above two claims mean
that $X_\alpha$ is a proper closed subset of $\GL(n)$ in the
Zariski topology whenever $\ket{\alpha}\neq 0$.

Similarly, if $\ket{\beta}\in U_{\lambda'}$ and
$\ket{v_{\lambda'}}$ is a highest weight vector for $U_{\lambda'}$
then $X_\beta := \{g\in\GL(n) |
\bra{v_{\lambda'}}g\ket{\beta}=0\}$ is a proper Zariski-closed
subset of $\GL(n)$ if and only if $\ket{\beta}\neq 0$.

The fact that $\ket{v_{\lambda}}$ and $\ket{v_{\lambda'}}$ are
highest weight vectors  means that
 $\ket{v_{\lambda}}\ot\ket{v_{\lambda'}}=\ket{v_{\lambda+\lambda'}}\in
U_{\lambda+\lambda'}$ and thus
\bea
\bra{v_\lambda}g\ket{\alpha}\bra{v_{\lambda'}}g\ket{\beta} &=&
(\bra{v_{\lambda}}\ot\bra{v_{\lambda'}})g
(\ket{\alpha}\ot\ket{\beta})\nonumber\\
 &=& \bra{v_{\lambda+\lambda'}} g
(\ket{\alpha}\circ\ket{\beta}). \label{eq:zero-div-RHS}
\eea
We are free to replace $\ket{\alpha}\ot\ket{\beta}$ with
$\ket{\alpha}\circ\ket{\beta}$ in the last step because we are
taking the inner product with a vector that lies entirely in
$U_{\lambda+\lambda'}$.  Now suppose
$\ket{\alpha}\circ\ket{\beta}=0$. Then for all $g$ at least one of
the terms on the LHS of \eq{zero-div-RHS} vanishes, and thus
$\GL(n)=X_\alpha \cup X_\beta$.  Since $\GL(n)$ is irreducible, it
cannot be the union of two proper closed subsets, and we conclude
$\ket{\alpha}$ and $\ket{\beta}$ cannot both be nonzero.

The proof of (b) is almost identical, but consider instead
$\ket{\alpha}\in U_\mu\ot U_\nu \ot U_\lambda^\star$,
$\ket{\beta}\in U_{\mu'}\ot U_{\nu'} \ot U_{\lambda'}^\star$ and
the group $\GL(m)\times \GL(n)\times \GL(mn)$ (which is still
irreducible).
\end{proof}

Note that we could relax the restriction that $\lambda_n\geq 0$ by
multiplying the inner products of the form
$\bra{v_\lambda}g\ket{\alpha}$ by a high enough power of $\det g$
(guaranteed to be nonzero for $g\in\GL(n)$) to obtain a polynomial
in the matrix elements of $g$.

\begin{proof}{\bf of theorem~\ref{theorem-stability}:}
Given any ring $R$ with an action of $G$ on it, let $R^G$ denote
the ring of $G$-invariants in $R$.  Now recall that if $|\mu|=m$
and $|\nu|=n$, then
\be g_{\mu\nu\lambda}= \dim (U_\mu \ot U_\nu \ot
U_\lambda^\star)^{\GL(m)\times \GL(n)},
\label{eq:g-invariant-Q}\ee where $\GL(m)\times \GL(n)$ acts on
$U_\lambda$ according to the inclusion
$\GL(m)\times \GL(n) \ra \GL(m)\otimes \GL(n) \subset \GL(mn)$.

If $g_{\mu \nu \lambda} \neq 0$ and $g_{\mu' \nu' \lambda'} \neq
0$ then according to \eq{g-invariant-Q} there exist nonzero
vectors $\ket{u_{\mu\nu\lambda}}\in(U_\mu \ot U_\nu \ot
U_\lambda^\star)^{\GL(m)\times \GL(n)}$ and
$\ket{u_{\mu'\nu'\lambda'}}\in(U_{\mu'} \ot U_{\nu'} \ot
U_{\lambda'}^\star)^{\GL(m)\times \GL(n)}$.  Define
$\ket{u_{\mu+\mu',\nu+\nu',\lambda+\lambda'}} =
\ket{u_{\mu,\nu,\lambda}} \circ\ket{u_{\mu',\nu',\lambda'}}$. Then
$\ket{u_{\mu+\mu',\nu+\nu',\lambda+\lambda'}} \neq 0$ by part (b)
of lemma~\ref{lemma:zero-div} and
$\ket{u_{\mu+\mu',\nu+\nu',\lambda+\lambda'}} \in U_{\mu+\mu'}\ot
U_{\nu+\nu'} \ot U_{\lambda+\lambda'}^\star$ is $\GL(m)\times
\GL(n)$-invariant, since this property is preserved by the Cartan
product. Thus $(U_{\mu+\mu'}\ot U_{\nu+\nu'} \ot
U_{\lambda+\lambda'}^\star)^{\GL(m)\times \GL(n)}\neq 0$ and we
conclude that $g_{\mu+\mu',\nu+\nu',\lambda+\lambda'} \neq 0$.
\end{proof}

Next, we prove that the semigroup $\KRON$ is finitely generated.

\begin{proof}{\bf of theorem~\ref{theorem-finitely-generated}:}
Suppose $V$ is a regular representation of a group $G$ and
$R=P(V)$ is the ring of polynomials on $V$. $G$ acts on $R$ by
$gp(v)=p(g^{-1}v)$, where $p$ is a polynomial function applied to
$v \in V$. One can therefore define $R^G$, the set of
$G$-invariant elements of $R$. In 1890, Hilbert proved that $R^G$
is finitely generated in the case where
$G=\SL(n,\complex)$~\cite{Hilbert1890a}. More generally, the
theorem holds whenever $G$ is {\em reductive}; this means that,
given a subspace $W$ of a representation $V$ that is closed under
the action of $G$, another subspace $W'$ can be found that is also
closed under $G$ and $V=W \oplus W'$. For example, GL(d) is
reductive. A short proof of this finite generation theorem can be
found in~\cite[Theorem 4.1.1]{GooWal98}.

To prove that $\KRON$ is finitely generated, first note that any
$\ket{\alpha} \in U_\lambda$ can be interpreted as the polynomial
$g \to \bra{v_\lambda}g\ket{\alpha}$, by the Borel-Weil theorem.
The Cartan product takes polynomials to products of polynomials,
as we have just seen (Eq.~\ref{eq:zero-div-RHS}). This ring is
generated by polynomials from the {\em fundamental
representations} $U^d_{\omega_i}$, where $\omega_i=(1^i)$ is the
diagram consisting of a single column of height $i$. This follows
because we can write the maximum weight vector $\ket{v_\lambda}$
for any $U_\lambda$ as the tensor product of maximum weight
vectors $\ket{v_{\omega_i}}$ of fundamental representations, just
by breaking down $\lambda$ into its columns of various heights.
There are, however, relationships amongst these generators; these
define an ideal $I$ closed under $G$.

Thus we can write $Q^d=R/I$, where $R=P(\bigoplus_i
U^d_{\omega_i})$ is the polynomial ring on the direct sum of the
fundamental representations. If we take $G=GL(d)$, since $G$ is
reductive and $I$ is closed under $G$, there is a subspace $J$ of
$R$, also closed under $G$, such that $R=I \oplus J$. Thus any
$G$-invariant of $R/I$ defines a $G$-invariant of $J$ and hence of
$R$. So $(R/I)^G=R^G/I^G$, and $(R/I)^G$ is finitely generated if
$R$ is. We now apply this argument to $R=Q^m \otimes Q^n \otimes
(Q^{mn})^\star$ with the reductive group $G=\GL(m) \times \GL(n)$
acting on it, and conclude that $R^G$ is finitely generated.

An invariant of $R^G$ is an element of $(U_\mu \otimes U_\nu
\otimes (U_\lambda)^\star)^{GL(m) \times GL(n)}$ for some triple
$(\mu, \nu, \lambda)$, and this defines an element of $\KRON$.
Since the Cartan product in R corresponds to the sum in $\KRON$,
the finite set of generators for $R^G$ define a finite set of
generators for $\KRON$.
\end{proof}

\section{The Correspondence of Nonzero Kronecker \newline Coefficients to Spectra}

\begin{proof}{\bf of theorem~\ref{ourtheorem}:}
Rather than working with the mixed state $\rho^{AB}$ we will
consider a purification $\ket{\psi}^{ABC}$ of $\rho^{AB}$, which
has $\spec \rho^C=\spec \rho^{AB}$. Let $r^{A}:=\spec \rho^{A},
r^{B}:=\spec \rho^{B}$ and $r^C:=\spec \rho^{C}$. $P^{A}_\mu$
denotes the projector onto the Young subspace $U_\mu \otimes
V_\nu$ in system $A$, and $P^B_\nu$, $P^C_\lambda$ are the
corresponding projectors onto Young subspaces in $B$ and $C$,
respectively. As a consequence of
theorem~\ref{theorem-Keyl-Werner} (see~\cite[Corollary
2]{ChrMit05}), for given $\epsilon>0$ one can find a $k_0$ such
that the following inequalities hold simultaneously for all $k
\geq k_0$:
\bea \label{equation-sum1} \tr P_X (\rho^A)^{\otimes k} &\geq&
1-\epsilon, \quad P_X:=\sum_{\mu: \|\bar{\mu} - r^A\|_1\leq\epsilon}
P^{A}_\mu\\
 \label{equation-sum2}\tr P_Y (\rho^B)^{\otimes k} &\geq& 1-\epsilon,
 \quad P_Y:=
\sum_{\nu: \|\bar{\nu} - r^B\|_1\leq\epsilon}
P^{B}_\nu\\
\label{equation-sum3} \tr P_Z (\rho^{C})^{\otimes k} &\geq&
1-\epsilon, \quad P_Z:=
\sum_{\lambda: \|\bar{\lambda} - r^C\|_1\leq\epsilon}
P^{C}_\lambda.
\eea
For $0< \epsilon< \frac{1}{3}$, the
estimates~(\ref{equation-sum1})-(\ref{equation-sum3}) can be
combined to yield
\be \label{equation-sum4} \tr [\left(P_X \otimes P_Y \otimes
  P_Z\right)
 (\proj{\psi}^{ABC})^{\otimes k}] \geq
1-3\epsilon>0.
\ee
Since $(\ket{\psi}^{ABC})^{\otimes k}$ is evidently invariant
under permutation of its $k$ subsystems, it takes the form
$$
(\ket{\psi}^{ABC})^{\otimes k}=\sum_{\mu \nu \lambda}
\ket{\alpha_{\mu \nu \lambda}},
$$
where $\ket{\alpha_{\mu \nu \lambda}} \in U_\mu \otimes U_\nu
\otimes U_\lambda \otimes (V_\mu \otimes V_\nu \otimes
V_\lambda)^{S_k}$. Equation~(\ref{equation-sum4}) then implies
that there must be at least one triple $(\mu, \nu, \lambda)$ with
$\|\bar{\mu} - r^A\|_1 \leq \epsilon$, $\|\bar{\nu} - r^B\|_1 \leq
\epsilon$, $\|\bar{\lambda} - r^C\|_1 \leq \epsilon$ and
$\ket{\alpha_{\mu \nu \lambda}} \neq 0$.  Thus $(V_\mu \ot V_\nu
\ot V_\lambda)^{S_k}\neq 0$ implies that $g_{\mu   \nu \lambda}
\neq 0$. It remains to pick a sequence of decreasing $\epsilon_j$
with corresponding triples $(\mu^{(j)}, \nu^{(j)},
\lambda^{(j)})$.
\end{proof}

It has been observed in different contexts that the speed of
convergence in theorem~\ref{theorem-Keyl-Werner} and consequently
in theorem~\ref{ourtheorem} is proportional to
$1/\sqrt{k}$~\cite{AlRuSa87, Harrow05}.

We will now prove corollary~\ref{cor-equiv}, the equivalence of
theorem~\ref{ourtheorem} and~\ref{theorem-Klyachko}.

\begin{proof}{\bf of corollary~\ref{cor-equiv}}
We start by showing how theorem~\ref{theorem-klyachko-our} follows
from theorem~\ref{ourtheorem}.

Let $(r^A, r^B, r^{AB}) \in \qmp$, the set of admissible spectral
triples. According to theorem~\ref{ourtheorem}, there is a
sequence of elements in $\KRON$, whose normalised values converge
to $(r^A, r^B, r^{AB})$. By theorems~\ref{theorem-stability}
and~\ref{theorem-finitely-generated}, the set $\KRON$ is a
finitely generated semigroup. With a finite set of generators
$(\mu^{(i)}, \nu^{(i)}, \lambda^{(i)})$ of $\KRON$ we can
therefore express $(r^A, r^B, r^{AB})$ in the form
\be \label{eq-vertex}
(r^A, r^B, r^{AB})= \sum_i x_i (\bar{\mu}^{(i)}, \bar{\nu}^{(i)},
\bar{\lambda}^{(i)}),
\ee
\sloppy for a set of nonnegative numbers $x_i$ which sum to one.
Since the union of the $t+1$-vertex simplices equals the whole
polytope, every point in it can be taken to be the sum of just
$t+1$ normalised generators $(\bar{\mu}^{(i)}, \bar{\nu}^{(i)},
\bar{\lambda}^{(i)})$ (cf.~Carath\'eodory's theorem). From the set
of $m+n+mn$ equations in the variables $x_i$ in
equation~(\ref{eq-vertex}), choose a set of $t$ linearly
independent ones, add the $(t+1)$'th constraint $\sum_i x_i=1$ and
write the set of equations as $ M\vec{x} = \vec{r},$
i.e.~$\vec{r}=(r_1, \ldots, r_t, 1)$ for $r_j \in \{ r^A_1,
\ldots, r^A_m, r^B_1, \ldots r^B_n, r^{AB}_1, \ldots,
r^{AB}_{mn}\}$ and $x=(x_1, \ldots, x_{t+1})$.

\fussy If $(r^A, r^B, r^{AB})$ is rational, the $x_i$ will be
rational as well, since $M$ is rational. This shows that $(r^A,
r^B, r^{AB})=\sum_i \frac{n_i}{n}(\bar{\mu}^{(i)},
\bar{\nu}^{(i)}, \bar{\lambda}^{(i)})$, where we set
$x_i=\frac{n_i}{n}$ for $n_i, n \in \naturals$. Multiplication by
$|\mu|n$ results in
    \bestar \label{eq-rational}
        |\mu|n(r^A, r^B, r^{AB})=\sum_i n_i({\mu}^{(i)}, {\nu}^{(i)},
{\lambda}^{(i)}).
    \eestar
Since the right hand side of this equation is certainly an element
of $\KRON$ this shows that for rational $(r^A, r^B, r^{AB})$ there
is a number $m:=|\mu|n$ such that $g_{m r^A, m r^B, m r^{AB}}\neq
0$.

It remains to show that theorem~\ref{ourtheorem} follows from
theorem~\ref{theorem-klyachko-our}. Suppose $(r^A, r^B, r^{AB})$
is a spectral triple corresponding to some $\rho^{AB}$. Then we
can construct a series of rational triples $(r^{A(j)}, r^{B(j)},
r^{AB(j)})$ that approaches $(r^A, r^B, r^{AB})$ and by
theorem~\ref{theorem-klyachko-our}, there exists a series
$(\mu^{(j)},\nu^{(j)},\lambda^{(j)})$ such that
$(\bar{\mu}^{(j)},\bar{\nu}^{(j)},\bar{\lambda}^{(j)})$ approaches
$(r^A, r^B, r^{AB})$ and
$g_{\mu^{(j)},\nu^{(j)},\lambda^{(j)}}\neq 0$ for all $j$.
\end{proof}

Note that there are two ways in which Klyachko's
theorem~\ref{theorem-klyachko-our} does not quite give the full
strength of theorem~\ref{ourtheorem}. First, it does not guarantee
the speed of convergence. Second, it does not imply that, in the
case of rational triples $(r^A, r^B, r^{AB})$, there is an
increasing sequence of values of $k$ for which $g_{kr^A, kr^B,
kr^{AB}} \ne 0$; this follows from theorem~\ref{ourtheorem} and
can be thought of as a sort of stability obtained without appeal
to theorem~\ref{theorem-stability}.

We will now turn out attention to theorem~\ref{theorem-converse}.
This theorem is a consequence of three things: compactness of the
set of density matrices on $\bbC^m\ot \bbC^n$, the semigroup
property (theorem~\ref{theorem-stability}) and the following
lemma, which may be thought of as a simpler and more quantitative
version of theorem~\ref{theorem-Klyachko}~\cite[Theorem
5.3.1]{Klyachko04}.

\begin{lemma}
Let $\mu, \nu$ and $\lambda$ be diagrams with $k$ boxes and at
most $m$, $n$ and $mn$ rows, respectively. If $g_{\mu \nu \lambda}
\neq 0$, then there exists a density operator $\rho^{AB}$ on
$\complex^{m} \otimes \complex^n$ with spectra
\bea \|\spec \rho^{A}-\bar{\mu}\|_1 &\leq& \delta
\label{eq:spec-approx-1}\\\label{eq:spec-approx-1a}
\|\spec \rho^{B}-\bar{\nu}\|_1 &\leq& \delta\\
\|\spec \rho^{AB}-\bar{\lambda}\|_1 &\leq& \delta
\label{eq:spec-approx-2}
\eea
for $\delta = O(mn\sqrt{(\log k)/k})$
\end{lemma}
Here if $p,q$ are probability distributions then $\|p-q\|_1 :=
\sum_x |p(x)-q(x)|$.

\begin{proof}
It will suffice to construct a pure state $\ket{\varphi}^{ABC} \in
\bbC^m \ot \bbC^n \ot \bbC^{mn}$ with $\|\spec
\varphi^A-\bar{\mu}\|_1\leq \delta$, $\|\spec
\varphi^B-\bar{\nu}\|_1\leq \delta$ and $\|\spec
\varphi^C-\bar{\lambda}\|_1\leq \delta$, since $\spec
\varphi^{AB}=\spec \varphi^C$.

By our assumption, there exists a unit vector
$\ket{\psi'}\in(V_\mu\ot V_\nu \ot V_\lambda)^{S_k}$.  Now extend
$\ket{\psi'}$ to a unit vector $\ket{\psi}\in (U_\mu\ot U_\nu \ot
U_\lambda)\ot (V_\mu\ot V_\nu \ot V_\lambda)^{S_k}$.  By
Schur-Weyl duality (Eq.~\ref{SWD}) we can embed this space in
$((\bbC^m \ot \bbC^n \ot \bbC^{mn})^{\ot k})^{S_k}$. Thus
$$\ket{\psi} \in
((\bbC^m \ot \bbC^n \ot \bbC^{mn})^{\ot k})^{S_k} \cong
((\bbC^{m^2n^2})^{\ot k})^{S_k}.$$ Again by Schur-Weyl duality,
$((\bbC^{m^2n^2})^{\ot k})^{S_k}$ is an irreducible representation
of $\GL(m^2n^2)$, which we denote by $U_{(k)}^{m^2n^2}$ to
emphasize which $\GL(\cdot)$ we are using.  Denote the projector
onto $U_{(k)}^{m^2n^2}\ot V_{(k)} \subset (\bbC^{m^2n^2})^{\ot k}$
by $P_{(k)}^{m^2n^2}$.  Note that $\tr P_{(k)}^{m^2n^2} = \dim
U_{(k)}^{m^2n^2} = \binom{k+m^2n^2-1}{m^2n^2-1}\leq k^{m^2n^2}$.
Fix a vector $\ket{\phi_0}\in\bbC^{m^2n^2}$ and let $dU$ denote a
Haar measure for $\U(m^2n^2)$ with normalisation $\int dU=1$. Then
by Schur's lemma
$$P_{(k)}^{m^2n^2} =
\dim U_{(k)}^{m^2n^2}  \int_{U \in \U(m^2n^2)} \!\!\!\! dU \;
(U \proj{\phi_0}U^\dagger)^{\otimes k}.$$ Thus
\beastar 1&=&\tr P_{(k)}^{m^2n^2} \proj{\psi} \\
 &=& \dim U_{(k)}^{m^2n^2} \int_{U \in \U(m^2n^2)} \!\!\!\! dU
\; \tr \proj{\psi} (U \proj{\phi_0}U^\dagger)^{\otimes k} \\
    &\leq& \dim U_{(k)}^{m^2n^2} \max_{U \in \U(m^2n^2)}
\; \tr \proj{\psi} (U \proj{\phi_0}U^\dagger)^{\otimes k}
\eeastar
Let $U\in \U(m^2n^2)$ be the unitary operator achieving the above
maximisation, and define $\ket{\varphi}:=U\ket{\phi_0}$.  Then
 $$|\bra{\psi}\left( \ket{\varphi}^{\otimes k}\right)|^2 \geq
\frac{1}{\dim U_{(k)}^{m^2n^2}} \geq k^{-m^2n^2}.$$
   Let $P_\mu^{m}$,
$P_\nu^{n}$ and $P_\lambda^{mn}$ denote the projectors onto
$U_\mu^m\ot V_\mu\subset (\bbC^m)^{\ot k}$, $U_\nu^m\ot
V_\nu\subset (\bbC^n)^{\ot k}$ and $U_\lambda^m\ot
V_\lambda\subset (\bbC^{mn})^{\ot k}$, respectively. Then by
construction $\left(P_\mu^{m}\ot P_\nu^{n} \ot
P_\lambda^{mn}\right)\ket{\psi} = \ket{\psi}$, so $\proj{\psi}
\leq P_\mu^{m}\ot P_\nu^{n} \ot P_\lambda^{mn}$ and
\beastar
\tr \left(P_\mu^{m}\ot P_\nu^{n}\ot P_\lambda^{mn}\right)
\proj{\varphi}^{\ot k} & \geq &
\tr (\proj{\psi}) \proj{\varphi}^{\ot k} \\
&=& |\bra{\psi}\left( \ket{\varphi}^{\otimes k}\right)|^2 \\
&\geq&\frac{1}{\dim U_{(k)}^{m^2n^2}} \geq k^{-m^2n^2}.
\eeastar
Focussing for now on the $A$ subsystem, we have
\be\tr P_\mu^{m}(\varphi^A)^{\ot k} \geq k^{-m^2n^2}.
\label{eq:conv-hi-overlap}\ee On the other hand, theorem~\ref{KW}
(Spectrum Estimation) states that
\be\tr P_\mu^{m}(\varphi^A)^{\ot k} \leq (k+1)^{{m(m-1)}/{2}}
\exp(-kD(\bar{\mu}\| \spec \varphi^A)).
\label{eq:conv-lo-overlap}\ee Combining
\eqns{conv-hi-overlap}{conv-lo-overlap}, we find that
$$D(\bar{\mu}\| \spec \varphi^A) \leq
\frac{\frac{1}{2}m(m-1)\log (k+1) + m^2n^2\log k}{k}$$ and for
$k>1$, we can apply Pinsker's inequality~\cite{Pinsker64} (which
states that $\|p-q\|_1^2/2 \leq D(p\|q)$ for any probability
distributions $p,q$) to bound
$$\|\bar{\mu}-\spec\varphi^A\|_1 \leq 3mn\sqrt{(\log k)/k}.$$
This proves eq.~(\ref{eq:spec-approx-1}).
Eqs.~(\ref{eq:spec-approx-1a}) and (\ref{eq:spec-approx-2}) follow
by repeating this argument (starting with \eq{conv-hi-overlap})
for $P_\nu^{n}$ and $P_\lambda^{mn}$.
\end{proof}

\section{Convexity}

\noindent Let us now gather together some implications of the
theorems. Let $\kron$ denote the normalised triples $(\bar\mu,
\bar\nu, \bar\lambda)$, where $(\mu, \nu, \lambda) \in \KRON$.
From theorem~\ref{ourtheorem} we know that any admissable spectral
triple, i.e. any point in \qmp, can be approximated by a sequence
in $\kron$ and therefore lies in $\overline \kron$, the closure of
$\kron$; thus $\qmp \subseteq \overline \kron$. From
theorem~\ref{theorem-converse} we know that $\kron \subseteq
\qmp$, and hence, since $\qmp$ is closed, $\overline \kron
\subseteq \qmp$. Thus we have
    $$ \qmp=\overline{\kron}.$$
Note that $\kron$ consists of rational points (normalised row
lengths of diagrams) and there are certainly operators with
irrational spectra. So $\kron$, unlike its closure, is a proper
subset of $\qmp$.

Theorems~\ref{theorem-stability}
and~\ref{theorem-finitely-generated} allow us to say more about
$\overline \kron$, and hence $\qmp$. The semigroup property of
$\KRON$ (theorem~\ref{theorem-stability}) implies that if
$(\bar{\mu}, \bar{\nu}, \bar{\lambda}), (\bar{\mu}', \bar{\nu}',
\bar{\lambda}') \in \kron$, then
$$(p\bar{\mu}+(1-p)\bar{\mu}', p\bar{\nu}+(1-p)\bar{\nu}',
p\bar{\lambda}+(1-p)\bar{\lambda}') \in \kron,$$ for every $p$
with $0 \le p \le 1$. Thus $\overline{\kron}$ is convex.
Furthermore, theorem~\ref{theorem-finitely-generated} implies that
there is a finite set of generators $(\mu^{(i)}, \nu^{(i)},
\lambda^{(i)})$ of $\KRON$, so any $(\bar\mu, \bar\nu,
\bar\lambda) \in \overline \kron$ can be written
$$(\bar\mu, \bar\nu,\bar \lambda)= \sum_i x_i
(\bar{\mu}^{(i)}, \bar{\nu}^{(i)}, \bar{\lambda}^{(i)}).$$
Thus $\overline \kron$ is a convex polytope. We enshrine this in
theorem~\ref{theorem-convex-polytope}.

An alternative proof for theorem~\ref{theorem-convex-polytope}
that makes use of Kirwan's convexity theorem for moment maps can
be found in~\cite[Chapter 2.3.6]{Christ05}.

\section{Acknowledgments}

The inspiration for this paper came from a discussion by one of us
(MC) with Allen Knutson, who essentially sketched the arguments of
theorems~\ref{theorem-stability}
and~\ref{theorem-finitely-generated}. He gave us further helpful
advice at several points, always worded in a lively and indeed
unforgettable way. We also had very valuable advice from Graeme
Segal. Finally, we thank Koenraad Audenaert for useful pointers to
the literature and Alexander Klyachko for many stimulating
discussions.

This project was supported by the EU under projects PROSECCO
(IST-2001-39227) of the IST-FET programme and RESQ
(IST-2001-37559). MC acknowledges the support of a DAAD
Doktorandenstipendium, the U.K.~Engineering and Physical Sciences
Research Council and a Nevile Research Fellowship, which he holds
at Magdalene College Cambridge. AWH thanks the Centre for Quantum
Computation for hospitality while completing this work and
acknowledges partial support from the ARO and ARDA under ARO
contract DAAD19-01-1-06.
\bibliographystyle{plain}

\end{document}

%% file: conversedef.tex
\usepackage{amsmath}
\usepackage{amssymb}
\usepackage{amsfonts}
\usepackage{theorem}
\usepackage{multicol}

\theoremstyle{break}
\newtheorem{theorem}{Theorem}[section]

\newtheorem{lemma}[theorem]{Lemma}

\newtheorem{corollary}[theorem]{Corollary}

\theoremstyle{plain}
\newtheorem{problem-footnote}[problem]{Problem}

\def\complex{\mathbb{C}}

\def\naturals{\mathbb{N}}

\def\01{\{0,1\}}

\newcommand{\ket}[1]{|#1\rangle}                    
\newcommand{\bra}[1]{\langle#1|}                    
\newcommand{\proj}[1]{|#1\rangle\langle#1|}         

\newcommand{\be}{\begin{equation}}
\newcommand{\ee}{\end{equation}}
\newcommand{\bea}{\begin{eqnarray}}
\newcommand{\eea}{\end{eqnarray}}
\newcommand{\bestar}{\begin{equation*}}
\newcommand{\eestar}{\end{equation*}}
\newcommand{\beastar}{\begin{eqnarray*}}
\newcommand{\eeastar}{\end{eqnarray*}}

\newcommand{\spec}{\text{Spec}\;}
\newcommand{\tr}{\text{Tr}\, }
\def\openone{\leavevmode\hbox{\small1\normalsize\kern-.33em1}}

\newcommand{\kron}{{\rm{KRON}}}                    
\newcommand{\KRON}{{\rm{Kron}}}                    
\newcommand{\qmp}{{\rm{SPEC}}}                     

\newcommand{\U}{{\rm U}}
\newcommand{\GL}{{\rm{GL}}}
\newcommand{\SL}{{\rm{SL}}}

\newcommand{\cH}{{\cal H}}

\newenvironment{proof}
{\noindent {\bf Proof }} {{\hfill $\Box$} \\
\noindent}

\newcommand{\captionfonts}{\footnotesize}

\makeatletter  
\long\def\@makecaption#1#2{%
  \vskip\abovecaptionskip
  \sbox\@tempboxa{{\captionfonts #1: #2}}%
  \ifdim \wd\@tempboxa >\hsize
    {\captionfonts #1: #2\par}
  \else
    \hbox to\hsize{\hfil\box\@tempboxa\hfil}%
  \fi
  \vskip\belowcaptionskip}
\makeatother   

\newlength{\quoteLength}
\newcounter{protoCount}
\newcounter{protoList}
\newsavebox{\tmpbox}
\newlength{\protobox}

\def\bbC{\mathbb{C}}
\def\bbZ{\mathbb{Z}}

\def\ot{\otimes}
\def\ra{\rightarrow}
\newcommand{\eq}[1]{eq.~(\ref{eq:#1})}
\newcommand{\eqns}[2]{eqns.~(\ref{eq:#1}) and (\ref{eq:#2})}
\DeclareMathOperator{\Par}{Par}